\documentclass[cameraready]{Interspeech}

\title{SISER: Speaker-Invariant Speech Emotion Recognition \\
with Entropy-Based Adversarial Training}

\author[ orcid={0009-0008-3205-8430}]{Eunseo}{Choi}
\author[ orcid={0009-0000-3470-5381}]{Hyunku}{Kang}
\author[ orcid={0000-0003-4085-2470}, correspondingauthor]{Chanwoo}{Kim}

\address{
 Korea Univiersity, South Korea
}

\email{\{ces4669, kahk000, chanwcom\}@korea.ac.kr}

\keywords{speech emotion recognition, Adversarial Training, Speaker Invariant}

\usepackage{comment}
\usepackage{graphicx}
\usepackage{subcaption}
\usepackage{tabularx}
\usepackage{multirow}
\usepackage{booktabs}
\usepackage{makecell}
\usepackage{microtype}
\newcommand{\diff}[1]{{\scriptsize(#1)}}

\begin{document}

\maketitle

\footnotetext{ The source code is publicly available at: \url{https://github.com/slp-lab-research/siser.git}.}

\begin{abstract}
    Speech emotion recognition (SER) faces two fundamental challenges: scarcity of labeled data and inter-speaker variability, both of which hinder generalization of emotion recognition systems. While prior adversarial approaches address speaker variability, they fall short in leveraging powerful pre-trained representations. We propose SISER (Speaker-Invariant Speech Emotion Recognition), integrating wav2vec~2.0 as a feature encoder and ECAPA-TDNN as a speaker discriminator within an entropy-based adversarial training scheme. wav2vec~2.0 provides rich self-supervised representations that alleviate dependency on large labeled datasets, while ECAPA-TDNN enables suppression of speaker identity via a stronger adversarial signal than shallow classifiers. Evaluated on IEMOCAP, SISER achieves a UA of 60.63\%, outperforming the baseline (51.15\%) and wav2vec~2.0 without speaker suppression (56.46\%), with ablation emphasizing that the choice of speaker classifier architecture is a key factor.
\end{abstract}

\section{Introduction}

Human speech carries a diverse mixture of information, including phonetic content, speaker characteristics, and affective state~\cite{schuller2018speech, 10446852}. These components are deeply intertwined at the acoustic level, making it difficult to isolate emotion-relevant information from speaker-specific patterns. In the context of speech emotion recognition(SER), speaker variability is widely recognized as one of the most persistent sources of performance degradation~\cite{schuller2013interspeech,zhan1997speaker,el2011survey}. This is particularly problematic in speaker-independent settings, where models trained on a fixed set of speakers tend to capture speaker-specific correlations that do not generalize to unseen speakers, a form of domain mismatch that directly undermines cross-speaker generalization. Furthermore, the scarcity of labeled emotion data poses an additional challenge, as collecting and annotating such data is costly and time-consuming~\cite{kratzwald2018deep, neumann2019improving}.

Self-supervised pretraining, as employed in wav2vec~2.0~\cite{baevski2020wav2vec}, offers a promising solution by learning transferable representations from large amounts of unlabeled speech, reducing dependency on costly annotated data. These representations have been shown to transfer effectively to emotion recognition tasks even under limited supervision~\cite{pepino2021emotion, wang2021fine}. However, even with such powerful pretrained representations, speaker information remains entangled with emotion content. Fine-tuning a pretrained encoder on an emotion classification objective alone does not guarantee that speaker-specific patterns are suppressed in the resulting embedding, which limits generalizability to unseen speakers.

Adversarial training has emerged as a principled approach to removing speaker information from learned representations~\cite{ganin2016domain}. In this framework, a speaker classifier is trained to identify the speaker from the encoder output, while the encoder is simultaneously trained to confuse it. Li et al.~\cite{li2020speaker} applied this strategy to SER, proposing an entropy maximization objective as a stronger alternative to gradient reversal. Nevertheless, the role of the speaker classifier architecture has received limited attention. We argue that a weak speaker classifier exerts insufficient adversarial pressure, leaving speaker-relevant information in the encoder output.

In this work, we propose Speaker-Invariant Speech Emotion Recognition(SISER), building upon with two key modifications. First, we replace the convolutional and recurrent feature encoder with wav2vec~2.0 for richer, more transferable representations. Second, we replace the shallow linear speaker classifier with ECAPA-TDNN~\cite{desplanques2020ecapa}, a state-of-the-art speaker verification model capable of detecting speaker-relevant patterns that a linear classifier would miss. By adversarially training against this stronger discriminator and combining it with an entropy maximization objective, the encoder is pressured to produce representations equally attributable to any speaker, effectively removing speaker-discriminative content.

The main contributions of this work are as follows:
\begin{itemize}
    \item We propose a speaker-invariant SER framework that combines wav2vec~2.0 and ECAPA-TDNN within an entropy-based adversarial training scheme.
    \item We demonstrate that a stronger speaker discriminator leads to more thorough disentanglement, and that the speaker classifier architecture is important as shown in table2 .
\end{itemize}

\section{Related Works}
\subsection{Self-supervised representations for speech emotion recognition}
Self-supervised pretraining has produced strong general-purpose speech representations that transfer effectively to downstream tasks~\cite{kangBSS2026, 10096854}.
wav2vec~2.0 processes raw waveform input and learns contextualized frame-level features by solving a contrastive prediction task over masked segments. 
Several studies have applied such representations to SER. 
wav2vec~2.0 features have been shown to outperform hand-crafted 
acoustic features under limited supervision~\cite{pepino2021emotion}, 
and fine-tuning pretrained speech models has been demonstrated 
to yield strong emotion recognition performance~\cite{wang2021fine}. 
The layer-wise properties of self-supervised representations 
have also been explored for affective tasks, revealing that 
different layers encode complementary acoustic information~\cite{morais2022speech}. These findings demonstrate 
that self-supervised representations can be effectively leveraged for emotion recognition even under limited supervision. In our work, we build on wav2vec~2.0 as the feature encoder and focus on further improving speaker independence through adversarial training.

\subsection{Adversarial training and speaker discriminator design}
Domain adversarial training (DAT) was originally proposed as a general framework for learning domain-invariant representations~\cite{ganin2016domain}, and has since been applied across a range of tasks including 
automatic speech recognition~\cite{shinohara2016adversarial, sun2017unsupervised, meng2018speaker}, speaker adaptation~\cite{wang2018unsupervised}, and SER to reduce the influence of nuisance factors such as speaker identity~\cite{abdelwahab2018domain, tu2019adversarial}.
Within this framework, gradient reversal has been widely adopted, but it only penalizes the encoder for being discriminative without constraining the target distribution, meaning that collapsing representations onto a single alternative domain can equally satisfy the objective. Entropy maximization addresses this by explicitly encouraging a uniform posterior over all source domains~\cite{li2020speaker,grandvalet2004entropy_min}. 

However, prior frameworks have predominantly used shallow classifiers as the speaker discriminator, without considering the impact of discriminator capacity on disentanglement quality~\cite{abdelwahab2018domain, 
li2020speaker}. ECAPA-TDNN 
incorporates multi-scale temporal context aggregation and 
channel-dependent attention to produce highly discriminative speaker embeddings, making it a strong candidate for the adversarial speaker discriminator role. To the best of our knowledge, our framework is the first to use ECAPA-TDNN as the adversarial speaker discriminator within an entropy-based adversarial training scheme for SER.

\begin{figure}[t]
    \centering
    \includegraphics[width=1.0\linewidth]{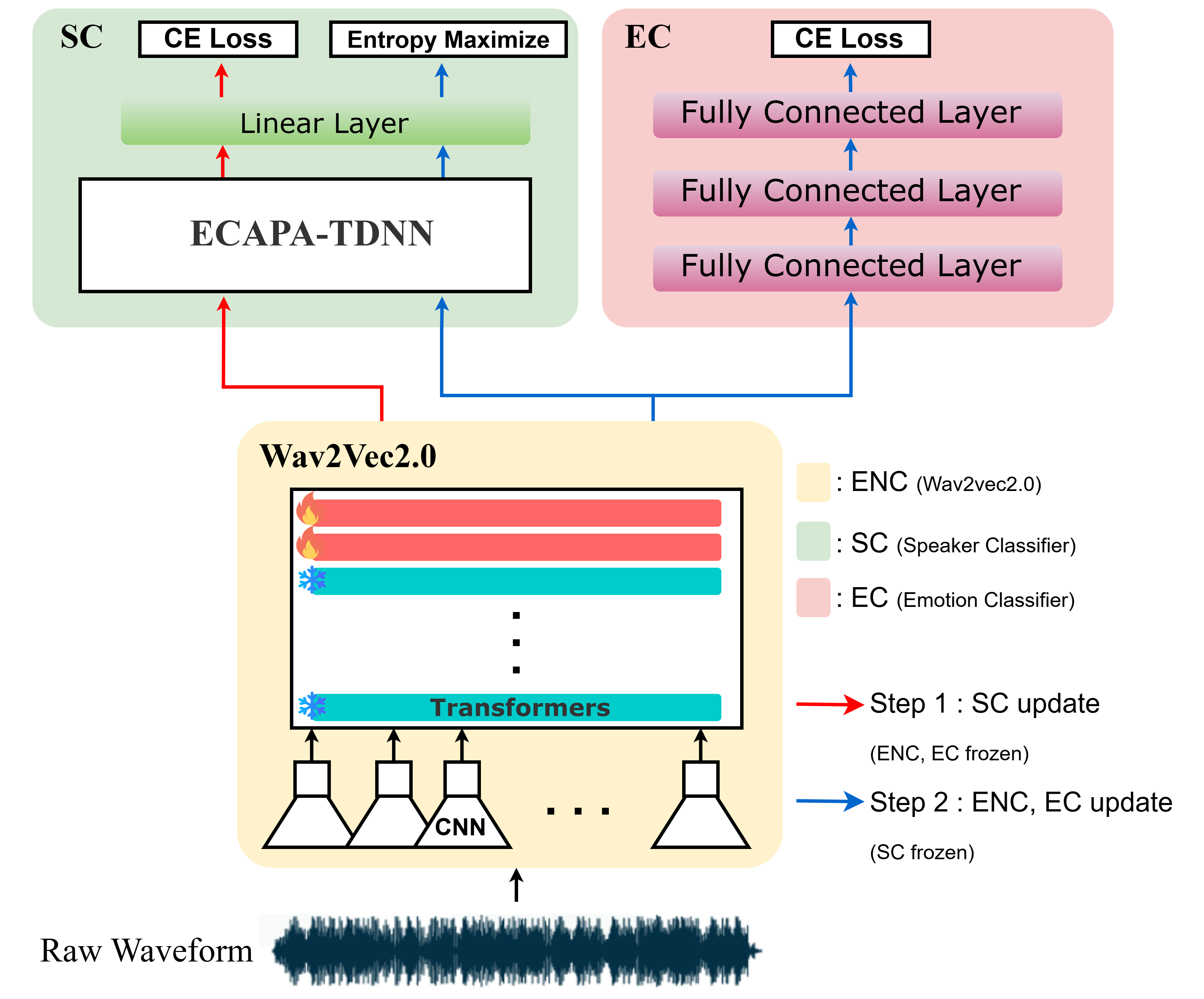}
    \caption{Overview of the proposed framework. Red and blue arrows denote the two alternating training steps: red for Step 1 (SC update with ENC and EC frozen) and blue for Step 2 (ENC and EC update with SC frozen).}
    \label{fig:model overall architecture}
\end{figure}

\begin{table*}[t]
\centering
\label{Main results}
\caption{Performance comparison on IEMOCAP under a 
speaker-independent 10-fold cross-validation protocol (UA/WA in \%). Five systems are compared: the baseline reproduced from~\cite{li2020speaker} with and without data augmentation (O/X), a wav2vec~2.0 encoder without speaker disentanglement (vanilla), and wav2vec~2.0 with ECAPA-TDNN trained via Gradient Reversal Layer (GRL) or entropy maximization (SISER). Results are reported on both validation and test sets. Best results per set are highlighted in bold.}
\resizebox{\textwidth}{!}{
\begin{tabular}{l c c c c c c c c}
\toprule
\multirow{2}{*}{Feature Extractor} &
\multirow{2}{*}{Data Aug} &
\multirow{2}{*}{Emotion Classifier} &
\multirow{2}{*}{Speaker Classifier} &
\multirow{2}{*}{SI Method} &
\multicolumn{2}{c}{Valid} &
\multicolumn{2}{c}{Test} \\

\cmidrule(lr){6-7} \cmidrule(lr){8-9}

 &  &  &   &  & UA(\%)$\uparrow$ & WA(\%)$\uparrow$ & UA(\%)$\uparrow$ & WA(\%)$\uparrow$ \\
\midrule
CNN + GRUs~\cite{li2020speaker}
& O
& FC 
& FC 
& Entropy 
& 60.24 & 58.85
& 59.91 & \textbf{\textcolor{blue}{58.62}} \\

CNN + GRUs (baseline)~\cite{li2020speaker}
& X
& FC 
& FC 
& Entropy 
& \textbf{54.51} & \textbf{54.20}
& \textbf{51.15} & \textbf{50.14} \\

\midrule

wav2vec~2.0 (vanila) 
& X
& FC 
& - 
& -
& 59.22 & 59.17 
& 56.46 & 54.45 \\

wav2vec~2.0 
& X
& FC 
& ECAPA-TDNN + Linear  
& GRL 
& 58.93 & 58.69 
& 56.95 & 56.01 \\

wav2vec~2.0 (ours) 
& X
& FC 
& ECAPA-TDNN + Linear  
& Entropy 
& \textbf{\textcolor{blue}{62.37}} & \textbf{\textcolor{blue}{61.64}} 
& \textbf{\textcolor{blue}{60.63}} & 58.53 \\
\bottomrule
\end{tabular}
}

\end{table*}

\section{Methodology}

Our goal is to obtain a fixed-dimensional utterance embedding in which emotion-related information is maximized while speaker-related information is removed. This is achieved through an adversarial training procedure between the feature encoder and a strong speaker classifier, described in detail below.

\subsection{Model Structure}

Our proposed model consists of three modules: the wav2vec~2.0 
feature encoder ENC, the emotion classifier EC, and the 
ECAPA-TDNN speaker classifier SC.

The ENC module is based on wav2vec~2.0, which processes raw waveform input and produces a sequence of frame-level contextualized representations. The resulting frame-level sequence is passed directly to the downstream modules as the utterance representation $v$, where $v = \text{enc}(x)$.

The emotion classifier EC consists of stacked fully 
connected layers that map $v$ to a distribution over emotion 
categories. The speaker classifier SC is implemented as 
ECAPA-TDNN, which uses squeeze-and-excitation residual blocks with multi-scale temporal context, followed by a linear output layer mapping to the number of training speakers in each fold.

\subsection{Adversarial Training}

Our training approach differs from prior adversarial SER 
frameworks along two dimensions. First, regarding the training 
objective, gradient reversal optimizes the encoder to merely 
perform poorly on speaker classification without specifying how 
it should fail, whereas entropy maximization explicitly targets 
a uniform output distribution over all speaker identities. 
Second, regarding the speaker classifier, prior methods rely on 
shallow fully connected layers whose limited capacity allows 
some speaker-relevant dimensions of the encoder output to evade 
detection, resulting in incomplete disentanglement. Our proposed 
framework addresses both limitations, as described below.

Training dataset $\mathcal{D}=\{(X_1,e_1,k_1), \ldots,(X_i,e_i,k_i),\ldots,\\
(X_N,e_N,k_N)\}$ contains $N$ samples in which each speech
utterance $X_i$ is produced by speaker $k_i$ with emotion label
$e_i$. $\mathcal{X}$, $\mathcal{E}$, and $\mathcal{K}$ denote
the sets of all utterances, emotion categories, and speaker
identities respectively.

\subsubsection{Training of SC}
SC serves as a discriminator that predicts speaker identity 
from the input representation $v$. The parameters of SC 
are updated by minimizing the cross-entropy loss over speaker 
identities:
\begin{equation}
    L_{D\_Spk} = -\vspace{-2mm}\sum\vspace{-2mm}_{(X_i, e_i, k_i) \in \mathcal{D}} 
    \log P(k_i \mid \text{enc}(X_i))
    \label{eq:spk}
\end{equation}\\
During this step, ENC and EC are kept frozen, and only SC is 
updated to improve speaker identification from the encoder 
output.

\subsubsection{Training of ENC and EC}
ENC and EC are jointly optimized to retain emotion-relevant 
content in the encoder output while eliminating 
speaker-discriminative information. To this end, the emotion 
classification loss minimizes cross-entropy over emotion 
categories as in Equation~(\ref{eq:emo}), ensuring that the 
representation remains useful for emotion prediction.
\begin{equation}
    L_{D\_Emo} = -\vspace{-2mm}\sum\vspace{-2mm}_{(X_i, e_i, k_i) \in \mathcal{D}} 
    \log P(e_i \mid \text{enc}(X_i))
    \label{eq:emo}
\end{equation}\\
We drive the encoder to remove speaker cues by maximizing the 
entropy of SC's output distribution across all training 
speakers, as defined in Equation~(\ref{eq:entropy}). A 
maximally uniform posterior over all speaker identities implies 
that no speaker can be reliably inferred from the learned 
representation.
\begin{equation}
\begin{aligned}
L_{H_{\text{Spk}}}
= -\vspace{-2mm}\sum_{(X_i, e_i, k_i) \in \mathcal{D}}
  \sum_{k_j \in \mathcal{K}}\vspace{-2mm}
   P(k_j \mid \text{enc}(X_i))
  \log P(k_j \mid \text{enc}(X_i))
\end{aligned}
\label{eq:entropy}
\end{equation}\\
The two losses are combined as in Equation~(\ref{eq:total}), 
where $\lambda \in (0, 1)$ balances emotion classification 
against speaker entropy maximization. A larger $\lambda$ 
emphasizes emotion classification, while a smaller $\lambda$ 
enforces stronger speaker suppression at the risk of discarding emotion-relevant information. SC is kept frozen during this 
step.
\begin{equation}
    \mathcal{L}(\theta_{\text{ENC}}, \theta_{\text{EC}}) = 
    \lambda \cdot L_{D\_Emo} - (1 - \lambda) \cdot L_{H\_Spk}
    \label{eq:total}
\end{equation}
As SC is realized as ECAPA-TDNN, the adversarial gradient 
flowing back to ENC captures speaker-relevant patterns across 
multiple temporal scales with channel-wise attention, yielding 
a richer training signal than a shallow classifier could 
provide. Consequently, ENC is pushed toward more complete 
removal of speaker-identifying cues under this training scheme.

\begin{figure*}[t]
    \centering

    \begin{subfigure}[t]{0.32\textwidth}
        \centering
        \includegraphics[width=\textwidth]{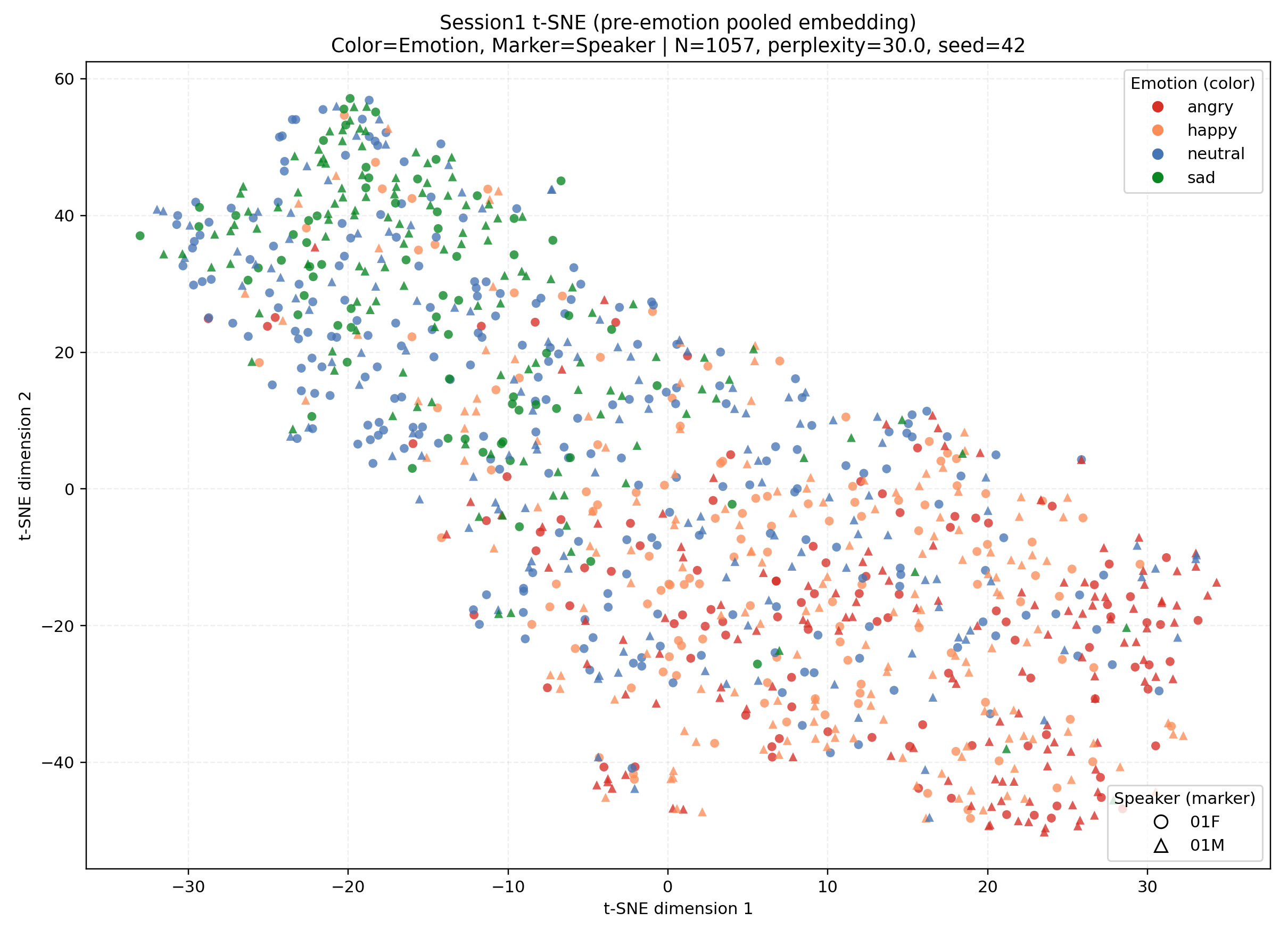}
        \caption{Baseline}
        \label{fig:tsne_Baseline}
    \end{subfigure}\hfill
    \begin{subfigure}[t]{0.32\textwidth}
        \centering
        \includegraphics[width=\textwidth]{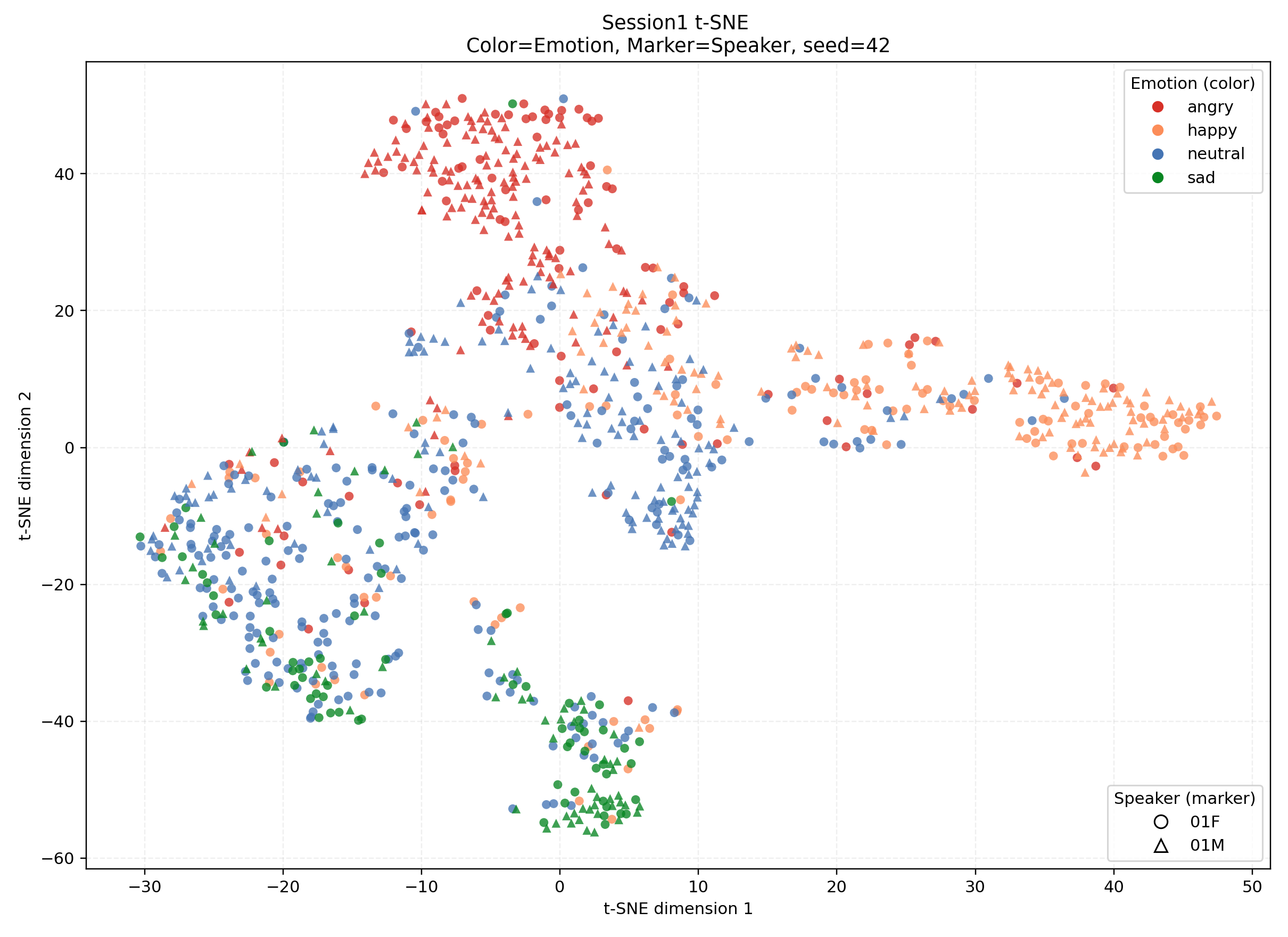}
        \caption{SISER}
        \label{fig:tsne_SISER}
    \end{subfigure}\hfill
    \begin{subfigure}[t]{0.32\textwidth}
        \centering
        \includegraphics[width=\textwidth]{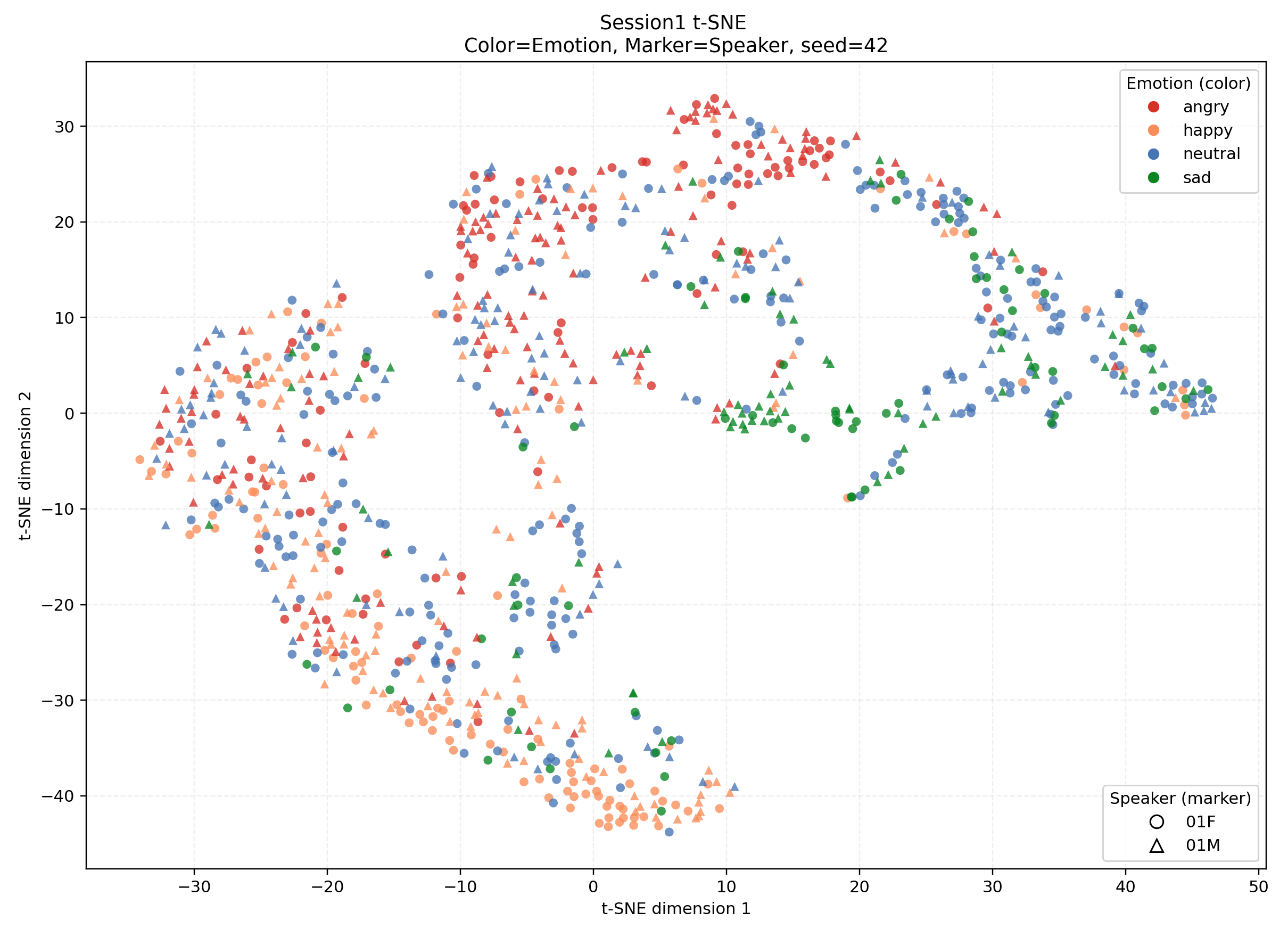}
        \caption{Vanilla}
        \label{fig:tnse_vanila}
    \end{subfigure}

    \caption{t-SNE visualizations of utterance-level embeddings for two speakers in Session 1. Color indicates emotion category and marker shape indicates speaker identity ($\circ$: 01F, $\triangle$: 01M).}
    \label{fig:tsne_visualization}
\end{figure*}

\section{Dataset}

\noindent We evaluate our proposed framework on the IEMOCAP dataset~\cite{busso2008iemocap}, which consists of five sessions of scripted and improvised dyadic conversations between pairs of actors. Each session involves one female and one male speaker, yielding ten speakers in total. The corpus provides categorical emotion annotations, and we follow the common practice of merging excitement and happiness to obtain a more balanced label distribution~\cite{yang2021superb}. This results in 5,531 utterances across four emotion classes, with 1,103 angry, 1,636 happy, 1,708 neutral, and 1,084 sad utterances.

Our baseline reproduces the method of \cite{li2020speaker, ko2015audioaug} without speed-perturbation data augmentation, which in the original study uses four different speed ratios to expand the training set to five times its original size. Our experiments are conducted on the original training samples without augmentation in order to isolate the contribution of the encoder architecture and speaker classifier design, and to reflect a more practical setting where augmentation may not always be applicable.

\section{Experiment Setup}
\noindent\textbf{Feature extraction.}
For the baseline system, we follow the feature extraction
procedure of~\cite{li2020speaker}. For the wav2vec~2.0 variants, the pretrained wav2vec~2.0 base model is used as the feature encoder and is fine-tuned during training on IEMOCAP, with only the last two transformer layers unfrozen~\cite{chang2023spin,chang2024rspin,chang2025sscrl}.

\noindent\textbf{General settings.}
Main experiments follow a 10-fold leave-one-session cross-validation protocol~\cite{cho2018deep,bpasts2020_136300,lee2020emotion}, with eight speakers for training, one for validation, and one for testing~\cite{li2020speaker}. All reported results reflect performance on unseen speakers. 

\noindent\textbf{Model configurations.}
Training uses the Adam optimizer with a learning rate of
$1 \times 10^{-4}$, a batch size of 64, and 300 epochs.
The weighting parameter $\lambda$ is set to $0.5$,
following~\cite{li2020speaker}. The emotion classifier consists of three fully connected layers with PReLU activations. The ECAPA-TDNN speaker classifier follows the standard architecture~\cite{desplanques2020ecapa, okabe2018attentive_stats_pooling}, with the input dimension adjusted to 768 to match the output dimension of wav2vec~2.0, and a linear output layer mapping to the number of training speakers per fold. All experiments were conducted on a single NVIDIA A100 GPU.

\section{Results and Conclusion}

\subsection{Evaluation on IEMOCAP}

Table~1 reports the 10-fold cross-validation results 
for four systems. All results are obtained on the validation set, which in all cases consists of an unseen speaker. Both unweighted accuracy (UA) and weighted accuracy (WA) ~\cite{ma2024emotion2vec} are reported.

The baseline system, which reproduces the entropy maximization method of \cite{li2020speaker} without data augmentation, achieves 51.15\% UA and 50.14\% WA. The gap relative to the originally reported results is consistent with the absence of speed-perturbation augmentation~\cite{ko2015audioaug}, which substantially increases the effective training set size in the original work.

Replacing the convolutional and recurrent encoder with wav2vec~2.0 while using only the emotion classifier, without any adversarial component, yields 56.46\% UA and 54.45\% WA. This improvement of 2.31\% in UA over the baseline demonstrates the substantial benefit of pretrained self-supervised representations, even before any speaker disentanglement is applied.

Introducing ECAPA-TDNN as the speaker classifier with a GRL achieves 56.95\% UA and 56.01\% WA, which is slightly lower than the wav2vec~2.0 vanilla system, consistent with the known limitation of gradient reversal that it does not constrain the target distribution~\cite{long2018cdan}.  Entropy maximization addresses this by explicitly targeting a uniform distribution over all speakers, which we argue leads to more thorough and stable disentanglement.

Our proposed framework, which combines wav2vec~2.0 with 
ECAPA-TDNN and entropy maximization, achieves 60.63\% UA and 58.53\% WA on the test set. This represents an improvement of 9.48\% in UA over the baseline without data augmentation, demonstrating the effectiveness of combining a strong pretrained encoder with a discriminative speaker classifier under an entropy maximization objective. Notably, our framework achieves comparable performance to the augmented baseline (59.91\% UA) without any data augmentation, suggesting that speaker-invariant representation learning can compensate for the absence of 
augmentation-based data expansion.

\subsection{t-SNE Visualization}

Figure~\ref{fig:tsne_visualization} presents t-SNE visualizations~\cite{vandermaaten2008tsne} of utterance-level embeddings for two speakers in Session~1, colored by emotion category and differentiated by speaker marker. The baseline embedding exhibits a largely linear distributional pattern with considerable overlap across emotion categories, indicating 
limited emotion discriminability. The vanilla wav2vec~2.0 embedding shows some improvement in emotion separability, reflecting the richer pretrained representations, but emotion categories remain substantially mixed. In contrast, the proposed SISER framework produces embeddings in which emotion categories are more clearly separated, providing visual evidence that the adversarial training with ECAPA-TDNN and entropy maximization leads to more emotion-discriminative representations.

\subsection{Ablation Study}

Table~2 presents an ablation study comparing two feature encoder 
types and two speaker classifier configurations. The study is 
conducted on Fold~2, which best approximates the 10-fold mean 
accuracy across all systems as shown in Table~\ref{tab:10-fold experiments results}.

First, replacing the three-layer fully connected speaker classifier with ECAPA-TDNN yields an improvement of 7.13\% UA for the CNN+GRU encoder 
and 6.49\% UA for the wav2vec~2.0 encoder. These consistent gains across both encoder types support our central claim that a stronger speaker discriminator provides a more effective adversarial signal and leads to more thorough removal of speaker information.

Second, replacing the CNN + GRU encoder with wav2vec~2.0 yields a gain 
of 1.73\% UA with the three-layer fully connected speaker classifier and 1.09\% UA with ECAPA-TDNN. The smaller relative gain from the encoder upgrade when ECAPA-TDNN is used suggests that a strong speaker discriminator partially compensates for differences in the initial feature representation. Together, these findings suggest that the speaker classifier architecture has at least as large an impact on disentanglement quality as the choice of feature encoder.

\begin{table}[!ht]
\caption{Ablation study on Fold~2 (UA in \%). Two encoder types 
(CNN+GRU vs. wav2vec~2.0) and two speaker classifier 
configurations (FC vs. ECAPA-TDNN) are compared. $\Delta$ 
values indicate the gain from replacing each component.}
    \centering
    \setlength{\textfloatsep}{8pt}
    \label{Ablation study}
    \resizebox{\columnwidth}{!}{
    \begin{tabular}{l c c c}
    \toprule
     & FC & ECAPA-TDNN & $\Delta$ (Ecapa gain) \\
    \midrule
    CNN+GRU  & 55.09 & 61.64 & +7.13 \\
    wav2vec~2.0  & 56.24 & \textbf{62.73} & +6.49 \\
    \midrule
    $\Delta$ (wav2vec~2.0 gain) & +1.73 & +1.09 &  \\
    \bottomrule
    \end{tabular}
    }
\end{table}

\begin{table}[!ht]
  \setlength{\textfloatsep}{8pt}
  \caption{10-fold experiment results}
  \label{tab:10-fold experiments results}
  \centering
  \setlength{\tabcolsep}{3pt}
  \begin{tabular}{c c c c @{\hspace{6pt}} c}
    \toprule
    \textbf{N-Fold} & \textbf{Baseline} & \textbf{Vanila} & \textbf{Ours} & \textbf{Total $\Delta$} \\
    \midrule
    Fold 1  & 49.11 \diff{-5.09} & 58.63 \diff{-0.59} & 61.14 \diff{-1.59} & 7.27 \\
    \textcolor{blue}{\textbf{Fold 2}}  & 55.09 \diff{+0.89} & 57.22 \diff{-2.00} & 63.05 \diff{+0.32} & \textcolor{blue}{\textbf{3.21}} \\
    Fold 3  & 58.81 \diff{+4.61} & 58.09 \diff{-1.13} & 55.90 \diff{-6.83} & 12.57 \\
    Fold 4  & 57.14 \diff{+2.94} & 62.08 \diff{+2.86} & 60.63 \diff{-2.10} & 7.90 \\
    Fold 5  & 50.19 \diff{-4.01} & 54.02 \diff{-5.20} & 55.73 \diff{-7.00} & 16.21 \\
    Fold 6  & 53.82 \diff{-0.38} & 57.71 \diff{-1.51} & 57.22 \diff{-5.51} & 7.40 \\
    Fold 7  & 53.35 \diff{-0.85} & 62.30 \diff{+3.08} & 61.24 \diff{-1.49} & 8.42 \\
    Fold 8  & 48.95 \diff{-5.25} & 58.57 \diff{-0.65} & 57.99 \diff{-4.74} & 10.64 \\
    Fold 9  & 57.56 \diff{+3.36} & 63.46 \diff{+4.24} & 62.29 \diff{-0.44} & 8.04 \\
    Fold 10 & 57.94 \diff{+3.74} & 60.12 \diff{+0.90} & 60.41 \diff{-2.32} & 6.96 \\
    \midrule
    Mean & 54.20 $\pm$ 3.55 & 59.22 $\pm$ 2.68 & 62.73 $\pm$ 1.72 & -- \\
    \bottomrule
  \end{tabular}
\end{table}

\subsection{10-fold Stability Analysis}
Table~\ref{tab:10-fold experiments results} reports the per-fold results for all three systems. 
Among all folds, Fold~2 most closely approximates the 10-fold 
mean UA for all three systems, and is therefore used as the 
representative fold for the ablation study in Table~\ref{Ablation study}. 
Our proposed framework achieves a mean UA of $62.73\% \pm 1.72$, 
compared to $54.20\% \pm 3.55$ for the baseline and 
$59.22\% \pm 2.68$ for the vanilla system. The lower variance 
across folds suggests that the proposed adversarial framework 
produces more consistent representations across different 
speaker pairings, indicating improved robustness to the specific 
composition of the training set.

\subsection{Conclusion}


We propose a speaker-invariant SER framework combining wav2vec~2.0 with ECAPA-TDNN under an entropy maximization objective. Evaluated on IEMOCAP, the proposed framework achieves 
62.73\% UA, outperforming the baseline by 9.48\% and surpassing the augmented baseline without data augmentation. Our results demonstrate that the discriminative capacity of the speaker classifier is a key factor in adversarial disentanglement. 
We intentionally adopt a simple classification head to isolate the contribution of speaker-invariant representation learning, and expect that more sophisticated classifiers will yield further gains.

\section{Acknowledgments}
This work was supported in part by: the National Research Foundation of Korea
(NRF) grant funded by the Korean government (MSIT) under Grant No. RS-2025-24535409;
 the Institute of Information \& Communications Technology Planning \&
Evaluation (IITP) grant funded by the Korean government (MSIT) under Grant No.
RS-2019-II190079 for the Artificial Intelligence Graduate School Program at
Korea University; the Institute of Information \& Communications Technology
Planning \& Evaluation (IITP) grant funded by the Korean government (MSIT) under
Grant No. RS-2025-02304828 for the Artificial Intelligence Star Fellowship
Support Program to Nurture the Best Talents; and  the Institute of
Information \& Communications Technology Planning \& Evaluation (IITP) grant
funded by the Korean government (MSIT) under Grant No. RS-2025-25442867.

\section{Generative AI Use Disclosure}
AI is used just for the editing.

\bibliographystyle{IEEEtran}
\bibliography{mybib}

\end{document}